\documentclass{aa}  
\usepackage{graphicx}
\usepackage{txfonts}
\usepackage{xcolor}

\usepackage{ulem}

\usepackage{hyperref}
\usepackage{url}

\usepackage{natbib}
\def\gsim{\lower.5ex\hbox{$\; \buildrel > \over \sim \;$}}
\def\lsim{\lower.5ex\hbox{$\; \buildrel < \over \sim \;$}}
\begin{document}

   \title{The Potential Contribution of Radiative Supernova Remnants to the Local Cosmic Ray Spectrum}
\titlerunning{Contribution of radiative SNRs to the CR spectrum}

   \author{P. Cristofari
          \inst{1}, 
          V. H. M. Phan\inst{1,2}, 
 S. Recchia\inst{3,4}
          \and
S. Gabici \inst{1,5,6}
          }

   \institute{Laboratoire d’étude de l’Univers et des phénomènes eXtrêmes, LUX, Observatoire de Paris, Université PSL, Sorbonne Université, CNRS, 92190 Meudon, France\\
              \email{pierre.cristofari@obspm.fr}
              \and Institute for Theoretical Particle Physics and Cosmology, RWTH Aachen University, Sommerfeldstr. 16, 52074 Aachen, Germany
            \and Istituto Nazionale di Astrofisica (INAF) – Osservatorio Astrofisico di Arcetri, Largo Enrico Fermi 5, 50125, Firenze, Italy
            \and (IFJ-PAN) Institute of Nuclear Physics Polish Academy of Sciences, PL-31342 Krakow, Poland
\and 
Universit\'e de Paris Cit\'e, CNRS, Astroparticule et Cosmologie,  F-75006 Paris, France
\and Istituto Nazionale di Astrofisica (INAF) - Osservatorio di Astrofisica e Scienza dello Spazio (OAS), via Gobetti 101, I-40129 Bologna, Italy
}
   \date{Received September 15, 1996; accepted March 16, 1997}

 
  \abstract
   {
Recent observations of very-high-energy gamma rays from massive stellar clusters and microquasars have demonstrated that these systems can efficiently accelerate particles to very high energies, renewing the debate on the dominant sources of Galactic cosmic rays (CRs) and casting further doubt on the role of supernova remnants (SNRs), long considered the prime candidates. In these astrophysical environments, particle acceleration is generally thought to be driven by diffusive shock acceleration (DSA). The question of the duration of the phase in which DSA  can  efficiently take place is therefore essential to clarify the contribution of the various candidate CR sources.}
   {We aim at clarifying the contribution of SNRs to the Galactic CR spectrum, and the expected efficiency of DSA, especially accounting for possible particle acceleration during part of the radiative phase.}
   {We compute the total CR proton content at typical SNRs from DSA and reaccelerated pre-existing CRs, and discuss the importance of these two mechanisms throughout the adiabatic and early radiative phases.}
   {We find that if DSA and reacceleration take place during a fraction of the radiative phase, the total content of particles energized at SNRs can quickly overcome the typical budget required to account for the bulk of CRs. This suggests that the typical instantaneous efficiency, often defined as the fraction of fluid ram pressure converted into CRs would need to remain below $\lesssim 1$\% for most of the SNR evolution. Moreover, the radiative phase can contribute to shaping the spectra injected from typical SNRs. We illustrate that the difference between SNRs from thermonuclear and core-collapse SNe could help account for the measured difference in the helium and proton of the local interstellar spectrum. }
   {}


   \maketitle
%

\section{Introduction}
Observations in the very-high-energy (VHE) gamma-ray domain have helped identify and characterize several astrophysical sites that could contribute to the local interstellar spectrum (LIS), including supernova remnants (SNRs)~\citep{ackermann2013,HESSSNR2018}, massive stellar clusters~\citep{abramowski2012,aharonian2019,aharonian2022}, and microquasars~\citep{LHAASO2025}.
For a long time, SNRs have been considered the preferred candidates~\citep{blasi2013,gabici2019}, but this scenario is facing several difficulties, such as their ability to accelerate particles up to the knee, to inject the precise expected cosmic ray (CR) spectra in the interstellar medium (ISM), or to account for the features and deviations in the  local interstellar spectrum (LIS) departing from the $E^{-2.7}$ slope canonically estimated below the knee~\citep[see e.g.,][for a short review]{cristofari2021}. 
These questions are directly connected to a fundamental problem that is the efficiency of diffusive shock acceleration (DSA), the first-order Fermi mechanism~\citep{axford1977,krymskii1977,bell1978,blandford1978} responsible for the energization of particles at SNR shocks.

The efficiency of DSA has been estimated through a variety of methods, one of the most direct estimates comes from the observations in the gamma-ray domain of SNRs, for which the hadronic gamma-ray luminosity from accelerated protons interacting with the ISM is proportional to the content of accelerated protons times the target density $L_{\gamma} \propto L_{\rm CR} \times n_0$. These observations typically lead to estimates of the total CR proton content to be  from $\sim$ 0.5\% to 10\% of the typical total explosion energy of supernovae (SNe) $10^{51}$ erg~\citep{abdo2010,ackermann2013}. 

The diffuse Galactic gamma-ray emission, spatially correlated with the interstellar gas, provides an estimate of the Galactic cosmic-ray energy density. Combined with the cosmic-ray residence time inferred from secondary-to-primary nuclei $\tau_{\rm esc} \sim (1-3)\times10^{7}\ {\rm yr}$, this yields an estimate of the Galactic cosmic-ray luminosity, $L_{\rm CR}\sim10^{41}\,\mathrm{erg s^{-1}}$. Assuming that SNRs are the dominant sources of Galactic CRs, the corresponding DSA efficiency is on the order of $\eta_{\rm CR} \sim L_{\rm CR}/(\nu_{\rm SN} E_{\rm SN})$, where $\nu_{\rm SN}$ is the Galactic SN rate and $E_{\rm SN}\sim10^{51}\,\text{erg}$ the typical explosion energy. For $\nu_{\rm SN}\sim 1.4$--$5.8$ per century~\citep{strom1994,tammann1994,smartt2009,adams2013,ma2025}, this yields $\eta_{\rm CR}\sim 5 -15\%$. 

Here $\eta_{\rm CR}$ corresponds to the efficiency of DSA, as the overall fraction of the total SN explosion energy converted into CRs. This should be distinguished from another quantity often used in discussions of DSA, namely the instantaneous acceleration efficiency, usually expressed as the fraction $\xi$ of the fluid ram pressure, $\rho v_{\rm sh}^2$, converted into CRs at a given time, or sometimes as the fraction of the incoming energy flux, $(1/2)\rho v_{\rm sh}^3$, with $\rho$ the density upstream and $v_{\rm sh}$ the shock velocity.

The vast majority of known SNRs emitting VHE gamma rays and thus clearly known to efficiently energizing particles, are in the free expansion (FE) or Sedov-Taylor (ST) phase of their evolution, and it is often assumed that efficient particle acceleration cannot take place at radiative shocks. However, the detection of VHE gamma rays from novae, reveals a clear example of radiative shocks efficiently accelerating non-thermal particles~\citep{HESS_RSophiuchi2022}, and motivates us to investigate the possible contribution of the radiative phase of SNRs to the CR spectrum, that has often been overlooked. 

Moreover, the missing SNR problem \citep{ball2023,tsalapatas2024,green2025} is closely connected to the duration of the radiative phase and, more generally, to the lifetime of SNRs. The current number of known SNRs is typically on the order of $\sim 300$, whereas a supernova rate of $\sim 3$/century combined with a typical SNR lifetime of $20$--$100\,\mathrm{kyr}$ would imply a Galactic population of $\gtrsim 600$--$3000$ SNRs. Recent observations with Square Kilometre Array Observatory (SKAO) precursors and pathfinders have identified large numbers of candidate SNRs that could potentially double the known Galactic population \citep{anderson2025}, alleviating part of the discrepancy while simultaneously raising questions about the role of these previously undetected SNRs. The recent discovery of very extended radiative SNRs in the LMC, with a relatively high surface brightness \citep{smeaton2025}, reinforces the need for a better understanding of the late stages of SNR evolution.

\section{Diffusive shock acceleration and reacceleration at supernova remnants}
\label{sec:DSA}
The acceleration of freshly injected particles from the thermal pool (usual DSA) and the reacceleration of pre-existing CRs (DSRA) are considered. 
DSA is accounted for under the usual assumption that a fraction $\xi$ of the fluid ram pressure $\rho v_{\rm sh}^2$ at the shock is converted into CRs, where $\rho$  is the density and $v_{\rm sh}$ is the shock velocity. 

At the shock, the spectrum of accelerated particles is $f_{\rm acc}(p,t)=A(t)(p/p_0)^{-\alpha}\exp[-p/p_{\rm max}(t)]$, where $\alpha=4$ in the test-particle limit, $p_0=mc$, with $m$ the proton mass, and $p_{\rm max}$ is the maximum momentum of accelerated particles. Assuming that the particle spectrum is normalized between $p_{\rm min}$, the minimum momentum of particles injected into the acceleration process, typically assumed to lie in the suprathermal tail of the downstream Maxwellian distribution, $p_{\rm min}=\xi_{\rm inj}p_{\rm th,2}$, where $p_{\rm th,2}=(2 m k_{\rm B}T_2)^{1/2}$ is the downstream thermal momentum and $\xi_{\rm inj}\sim2$--$4$~\citep[see e.g.,][]{blasi2005}, and $p_{\rm max}$, the normalization is $A=(4-\alpha)\xi\rho v_{\rm sh}^{2}
\left[cp_0^\alpha\left(p_{\rm max}^{4-\alpha}-p_{\rm min}^{4-\alpha}\right)\right]^{-1}$
for $\alpha\neq4$, and $A=\xi\rho v_{\rm sh}^{2} \left[cp_0^4\ln\left(p_{\rm max}/p_{\rm min}\right)\right]^{-1}$ for $\alpha=4$.

The content of reaccelerated CRs at the shock is estimated as in~\citet{blasi2004,blasi2017}, where the pre-existing CRs are treated as boundary condition upstream $f_{\infty}$, and 
\begin{equation}
f_{\rm reac}(p,t)= \alpha
\int_{p_{0}}^{p} \frac{\mathrm{d}p'}{p'}\, \left( \frac{p'}{p}\right)^{\alpha}\, f_{\infty}(p')  \exp[-p/p_{\rm max}(t)]
\end{equation}
The pre-existing CRs are assumed to be the ones described by the LIS~\citep{bisschoff2019} as inferred from Voyager~\citep{cummings2016}, PAMELA~\citep{adriani2011,adriani2011b} and AMS-02 measurements~\citep{aguilar2015,aguilar2019}. 

$f_{\rm acc}$ and $f_{\rm reac}$ are the analytical results of the usual 1-dimensional plane shock. 
The main difference between the two terms is that an efficiency factor $\xi$ appears in $f_{\rm acc}$ while not in $f_{\rm reac}$ as the reacceleration corresponds to re-energization of all particles at the upstream boundary. At young SNRs currently detected in the gamma-ray domain (typically less than a few kyears old), $f_{\rm acc}$ usually dominates over $f_{\rm reac}$, although it has been shown that in specific situation, such as a low density, the contribution of reaccelerated electrons could be somewhat close to the contribution of particles accelerated from the thermal pool~\citep{cristofari2019}. Since the seed pre-existing CR spectrum is steeper than $p^{-4}$, then for $p> p_0$,  $f_{\rm reac} (p) \propto f_{\rm acc}  (p) \propto p^{-\alpha}$.

The total number of particles is estimated using an approach similar to that of~\citet{cristofari2021}. After energization at the shock, the fate of accelerated and reaccelerated particles can be treated in a similar way: they can either remain trapped inside the SNR ('trap'), where they suffer losses, or, for particles at the highest momenta, escape directly into the ISM ('esc').
The number of  particles $\textrm{d}n_{\rm trap}(p,t)$ of momentum $p$ around momentum $\text{d}p$, in a given time interval $\textrm{d}t$ reads: 
\begin{equation}
\text{d}n_{\rm trap}(p,t) = \textrm{d}t \frac{4 \pi}{r} r_{\rm sh}^2(t)   v_{\rm sh}(t)  \left[ f_{\rm acc}(p,t)+ f_{\rm reac}(p,t) \right] 4\pi p^{2} \text{d}p
\label{eq:dnacc}
\end{equation}
where $r$ is the compression factor at the shock.

The number of trapped particles with momentum $N_{\rm trap}$ at the end of the SN evolution ($t = T_{\rm end}$) results from the cumulative contribution of particles injected at earlier times with initial momentum $p' > p$, whose momentum has subsequently decreased to $p$ by $t = T_{\rm end}$:

\begin{equation}
\label{eq:number}
N_{\rm trap}(p) = \int_{t_0}^{T_{\rm end}}  \textrm{d}t \frac{4 \pi}{r} r_{\rm sh}^2(t) \;  v_{\rm sh}(t) \left(\frac{p'}{p}\right)^2\; \left[f_{\rm acc}(p',t)+f_{\rm reac}(p',t) \right] \frac{\text{d}p'}{\text{d}p}. 
\end{equation}

Adiabatic losses can be accounted for by following the momentum evolution of a particle injected at time $t'$ with initial momentum $p'$: 
\begin{equation}
\frac{\mathrm{d}p}{\mathrm{d}t}= - \frac{p}{\mathcal{L}} \frac{\mathrm{d} \mathcal{L}}{\mathrm{d}t}
\label{eq:change}
\end{equation}

where $\mathcal{L}$ is related to the change in volume between $t'$ and $t$,
\begin{equation}
\mathcal{L}(t,t') = \left( \frac{\rho_{\rm down}(t)}{\rho_{\rm down}(t')} \right)^{1/3},
\label{eq:conservation}
\end{equation}
 $\rho_{\rm down}$ is the downstream density. For adiabatic expansion one has 
$\rho_{\rm down} \propto P^{1/\gamma} \propto (\rho v_{\rm sh}^2)^{1/\gamma}$.~\citep[see e.g., discussions in][]{cristofari2021,morlino2021} 
Synchrotron losses are negligible for protons.

\subsection{Injection of particles from the thermal pool in DSA}

The efficiency of DSA is linked to the ability of particles to enter the acceleration process which depends on the shock geometry, magnetic turbulence, and plasma conditions. For example, the obliquity of the magnetic field with respect to the shock normal is expected to play a key role, with quasi-parallel shocks generally favoring ion acceleration, while quasi-perpendicular configurations tend to suppress particle acceleration (unless efficient cross-field diffusion)\citep{jokipii1987,caprioli2014a}. This efficiency is further regulated by the amplification of magnetic fields through  CR driven instabilities, which enhances particle confinement and increases the maximum attainable energies \citep{bell2004}. 

A critical aspect of this problem is the injection of particles from the thermal pool into the acceleration process. Despite decades of theoretical and numerical efforts, the question of how particles get injected into DSA from the thermal pool remains one of the major open issues in the theory of collisionless shocks \citep{jones1994,blasi2005,ellison2005,caprioli2010,battarbee2013}. In kinetic simulations (particle-in-cell or hybrid approaches), particle injection emerges from the shock microphysics: a fraction of the downstream thermal particles, typically in the suprathermal tail of the Maxwellian, is reflected at the shock front and undergoes pre-acceleration, for instance through shock drift acceleration and interactions with self-generated electromagnetic turbulence (e.g., Alfvénic or whistler waves, or else), eventually reaching energies sufficient to diffuse upstream and start performing  Fermi cycles \citep{riquelme2011,caprioli2014a,caprioli2014b}.

However, simulations remain so far subject to important limitations: e.g.,  most studies are performed in reduced dimensionality (one- or two-dimensional setups), which can artificially constrain the development of turbulence and particle transport. Full 3D simulations still remain computationally expensive and are typically restricted to limited spatial and temporal scales. Moreover, the separation between kinetic and astrophysical scales implies that simulations cannot capture the full evolution of  SNRS, and often rely on reduced ion-to-electron mass ratios or simplified upstream conditions. Therefore, making the connection between  injection efficiencies  that are measured in kinetic simulations to macroscopic models of DSA in astrophysical environments remains an open challenge \citep{caprioli2014b,park2015,ha2018}.

In contrast, semi-analytical models of DSA cannot capture these kinetic processes and instead adopt phenomenological injection recipes. For example, injected can be modeled via a thermal leakage mechanism, whereby only particles with momentum above a threshold $p_{\mathrm{inj}} = \zeta p_{\mathrm{th}}$ are injected into the acceleration process, with $\zeta \sim 3\text{-}4$ \citep{blasi2004,blasi2005}.

Observations of SNR shocks across the radio to VHE gamma-ray domain, along with numerical simulations, suggest that the instantenous efficiency $\xi$  likely decreases as the shock weakens~\citep{caprioli2014a,ha2018}. Owing to the poorly constrained temporal evolution of $\xi$, it is often assumed to be constant in time. 
In a more general framework, one may allow for a time-dependent injection parameter, $\xi(t)$, to account for the decreasing efficiency of particle injection as the shock weakens. Motivated by kinetic simulations indicating a strong dependence of injection on the shock Mach number, $\xi$ may be expected to decrease as the shock velocity decreases, for example following a scaling $\xi(t) \propto v_{\rm sh}(t)^{\beta}$, with typically $\beta=1-2$. For illustrative purposes we consider in the following a time-dependent case $\xi(t)$ where $\beta=1$ and $\xi(1 \text{kyear})=0.02$. It is worth noting that, in the case of DSRA, the process is expected to be independent of the injection efficiency, since it directly acts on a pre-existing population of energetic paricles, which can easily  undergo shock crossings and start performing Fermi cycles~\citep{blasi2004,blasi2017}. Hybrid simulations have confirmed that pre-existing CRs can be efficiently re-accelerated at shocks even when the fresh injection of thermal particles in DSA is inefficient \citep{caprioli2018}.

\subsection{The radiative phase}
\label{sec:rad}
In the conventional picture, DSA is not efficient in the radiative phase, due to the fact that the shock becomes weaker and strongly compressed, leading to a reduced Mach number and modified shock structure, while enhanced radiative cooling and ion-neutral damping suppress magnetic turbulence, thereby limiting particle acceleration and confinement~\citep{drury1996,bykov2000,morlino2016}. Clearly, the acceleration in the radiative phase up to the highest energies, e.g. the PeV range is not realistic, if only due  to their reduced velocity and strong radiative losses. Still, radiative SNRs  may still contribute to the production and re-acceleration of particles at lower energies, in particular in the GeV–TeV range. As a result, the radiative phase could play a non-negligible role in shaping the CR spectrum below the TeV range, even if its contribution to the highest-energy Galactic CRs is expected to be negligible.However, the possible role of the radiative phase in particle acceleration has recently received renewed attention~\citep{diesing2024,petruk2025}. Observations of evolved SNRs interacting with dense environments, such as Kes~17, also provide evidence for significant non-thermal particle populations at relatively late evolutionary stages~\citep{supan2023}.
Previous studies have also highlighted the role of the radiative phase in shaping the CR spectrum. For instance, \citet{zirakashvili2022} have included radiative cooling, photoionization, and ion-neutral damping, and found that enhanced acceleration and rapid escape can produce a hardening of the released proton and electron spectra in the 100-500 GeV range. 

In this work, we followed the approach of~\citet{cristofari2025}, and additionally took into account the possible effects of the cooling on the upstream medium, in order to investigate how this could shape the combined contribution of freshly accelerated and reaccelerated protons.
In the radiative phase, the shock structure departs significantly from the ideal adiabatic Rankine-Hugoniot conditions of strong shocks, due to the development of a cooling layer downstream and potentially, of a radiative precursor upstream. As the post-shock plasma cools efficiently, a substantial fraction of the incoming kinetic energy flux, $F_{\rm kin} = \tfrac{1}{2}\rho v_{\rm sh}^3$, is converted into radiation. This is often found in radiative shock models, in which the thermal energy generated at the shock is radiated as the downstream gas cools \citep{binette1985,dopita1995,dopita1996}.
We parametrize this radiative efficiency as $F_{\rm rad}=\eta_{\rm rad}F_{\rm kin}$, with $\eta_{\rm rad}\simeq\min(1,t_{\rm adv}/t_{\rm cool})$, where $t_{\rm adv}=R_{\rm sh}/v_{\rm sh}$ is the characteristic advection/dynamical timescale and $t_{\rm cool}=3k_{\rm B}T_2/[2n_{2}\Lambda(T_2)]$ is the post-shock radiative cooling time, with $n_2=4n_0$, $T_2$ the immediate post-shock temperature, and $\Lambda(T)$ the optically thin radiative cooling function~\citep{chevalier1994,draine2011}.

A fraction of this radiated energy, predominantly in ionizing photons, is absorbed upstream and produces a photoionization/heating precursor. 
Such photoionization precursors seem to be a generic feature of fast radiative shocks, in which UV photons emitted in the cooling of the post-shock gas propagate upstream and ionize and heat the incoming material ahead of the discontinuity~\citep{dopita1996,sutherland2017}. 
The corresponding temperature increase is estimated from the absorbed flux and results in an effective upstream temperature $T_1^\ast = T_{\rm ISM} + \Delta T$, where $\Delta T$ scales with the ratio of absorbed radiative flux to mass flux. This preheating reduces the effective Mach number of the flow entering the discontinuity, $\mathcal{M}_1^\ast = v_{\rm sh}/c_s(T_1^\ast)$, and thus weakens the hydrodynamic jump.

The ionization state of the upstream medium was treated in a simplified flux-based manner. Here we assumed that only a fraction of the radiated flux contributes to photoionization, parameterized by the velocity-dependent factor $f_{\rm ion}(v_{\rm sh})$ that reflects the increasing ionizing efficiency with shock velocity \citep{dopita1996}. The absorbed ionizing flux is then written as $F_{\rm ion}^{\rm abs} = f_{\rm abs}\, f_{\rm ion}(v_{\rm sh})\, F_{\rm rad}$, where $f_{\rm abs}$ represents the effective upstream absorption fraction. Rather than solving the full radiative transfer and ionization equations, we approximate the heating per each absorbed photon by a fixed excess energy $E_{\rm heat}=h\nu-\chi$, where $\chi$ is the hydrogen ionization potential. 

Our treatment is intentionally phenomenological and is designed to capture the main effects found in more sophisticated radiation-hydrodynamic shock calculations, such as those performed with the Paris-Durham code~\citep{godard2019,lehmann2022,godard2024a,godard2024b}. In particular, we mimic the heating and ionization of the upstream gas by the shock self-radiation, which controls the thermal and ionization structure of the radiative precursor~\citep{dopita1996,sutherland2017}.

In our approach, we implicity assume that ionization and recombination timescales remain short compared to the advection time through the precursor, so that the upstream medium reaches a quasi-steady ionization state that is mainly set by the local radiation field~\citep{gnat2007}. Observational evidence for such preheated and partially ionized gas ahead of shocks has been suggested by observations of SNRs, which revealed precursor temperatures up to $\sim 10^4 -10^5$~K~\citep{raymond2023}.

The shock structure can therefore be described as a radiative precursor followed by a residual hydrodynamic discontinuity and a downstream radiative cooling layer. The radiation emitted by the cooling gas heats and partially ionizes the upstream medium, yielding an effective preshock temperature $T_1^\ast=T_{\rm ISM}+\Delta T$ and a corresponding effective Mach number
$\mathcal{M}_1^\ast=\frac{v_{\rm sh}}{c_s(T_1^\ast)}$; 
In our phenomenological treatment, the compression ratio of the residual hydrodynamic jump is obtained from the Rankine--Hugoniot relations evaluated using $\mathcal{M}_1^\ast$,
$r_{\rm sub}=\frac{(\gamma+1)\mathcal{M}_1^{\ast2}}{(\gamma-1)\mathcal{M}_1^{\ast2}+2}.$
When the precursor heating becomes important, the effective Mach number decreases and the residual subshock may become weaker than the canonical strong-shock limit ($r_{\rm sub}<4$). The downstream radiative cooling layer is, however, can still provide additional compression, so that the total compression ratio across the complete radiative structure can still exceed four. This prescription is intended as a phenomenological parameterization of the possible feedback of the radiative precursor on the shock structure, rather than a fully self-consistent treatment of the coupling between radiation and the HD structure. Similar reductions of the residual hydrodynamic jump have been found in detailed multifluid shock calculations when an extended precursor develops, for example in continuous or Jump-type (CJ-type) shocks computed with the Paris-Durham code (where the magnetic pressure and ion-neutral precursor are usually reponsible for the shock modification)\citep{godard2019,lehmann2022,godard2024a,godard2024b}.
The appearance of a subshock embedded in a broader precursor is a well-established feature of nonlinear CR modified shocks~\citep{drury1981,malkov2001}. The detailed physics at an upstream precursor remains an important source of uncertainty: radiative pre-heating and ionization, together with CR-induced instabilities, may alter the thermodynamic and magnetic conditions encountered by the shock thus altering the development of turbulence and in turn the efficiency of particle acceleration~\citep{drury1986,wagner2007,capanema2026}.

\subsection{Ion-neutral damping and maximum energy of accelerated particles}
The period of the radiative phase that is of interest to us is in fact the one during which DSA (or DSRA) can be efficient. In the radiative phase, the ion-neutral damping, i.e., the damping of Alfvén waves due to the friction of ions and neutral atoms, thereby suppressing the turbulence needed to confine and accelerate charged particles,  has been shown to play an important role. 
Downstream, the post-shock cooling leads to a partially ionized medium with a large neutral fraction, and thus ion-neutral damping becomes crucial. However, for DSA, what is most important is what happens upstream. 
Radiative shocks are often pictured to propagate into dense, partially ionized media (upstream), and the low shock velocities reduce the growth rate of the streaming instability, making it easier for ion–neutral collisions to dominate and suppress Alfvén waves~\citep{draine1993,drury1996,bykov2000}.

Different approaches to ion- neutral damping limited acceleration allow to estimate the maximum energy of accelerated particle in this regime. A linear approach consists in balancing the growth rate of Alfvèn waves $\Gamma_{\rm CR}$ to the damping rate due to ion-neutral friction $\Gamma_{\rm IN}$ at the resonant scale $k \sim 1/r_{\rm L}(p)$: $\Gamma_{\rm CR}(k)=\Gamma_{\rm IN}(k)$. 
The growth rate of the resonant Alfvèn waves driven by the CR streaming at a wavenumber $k$ is given by:
\begin{equation}
\Gamma_{\rm CR}(k)
=
\frac{\pi}{4}
\frac{v_A}{W(k)}
\left[ p^4 v(p) \left|\frac{\partial f}{\partial x}\right| \right]_{p = p_{\rm res}},
\end{equation}~\citep{kulsrud1969}
where $v_A$ is the Alfvèn speed, $W(k)$ is the wave energy density per unit logarithmic bandwidth, $f$ is the CR distribution function, and the expression is evaluated at the resonant momentum $p_{\rm res}$. 
This expression highlights that the spatial gradient of CRs at the resonant momentum is important; It is thus convenient to rewrite it in terms of the CR pressure per logarithmic momentum interval: 
\begin{equation}
P_{\ln p}(p) \equiv \frac{dP_{\rm CR}}{d\ln p}  \propto\; p^4 v(p) f(p),
\end{equation}
so that the growth rate becomes
\begin{equation}
\Gamma_{\rm CR}(k) \propto \frac{v_A}{W(k)} \left| \frac{\partial P_{\ln p}}{\partial x} \right|_{p = p_{\rm res}}.
\end{equation}

Physically, this means that waves at scale $k$ resonate only with CRs of momentum $p_{\rm res}$, so that only a narrow logarithmic interval in momentum contributes to the growth of instabilites. 
For a power-law distribution close to $f(p) \propto p^{-4}$, the CR pressure is distributed approximately evenly per logarithmic interval, so that
\begin{equation}
\frac{dP_{\rm CR}}{d\ln p} \sim \frac{P_{\rm CR}}{\ln\Lambda},
\end{equation}
where $\ln\Lambda = \ln(p_{\max}/p_{\min})$. In this case, a simple estimate of the growth rate is
\begin{equation}
\Gamma_{\rm CR}
\sim
\frac{v_A}{W(k)}
\frac{P_{\rm CR}}{L\,\ln\Lambda},
\end{equation}
where $L$ is the characteristic length scale of the CR precursor.

The damping rate due to ion-neutral collisions is (from the two-fluid (ions and neutrals) dispersion relation~\citep[see e.g.,][]{recchia2022}
\begin{equation}
\Gamma_{\rm IN} \simeq \frac{\nu_{\rm IN}}{2},
\end{equation}
with
\begin{equation}
\nu_{\rm IN} \simeq  \delta  T_4^{0.4}\, n_n \; {\rm s^{-1}}.
\end{equation}, with $\delta \approx 1 - 8.4 \times 10^{-9}$~\citep{draine1993,drury1996,recchia2022}. 

To obtain a simple scaling, the precursor size can be estimated as
\begin{equation}
L \sim \frac{D(p)}{v_{\rm sh}},
\end{equation}
where $v_{\rm sh}$ is the shock velocity and $D(p)$ the diffusion coefficient. Assuming Bohm diffusion 
$D(p) \sim \frac{1}{3} r_L(p) v(p)$, so that $L \propto p/(B v_{\rm sh})$,
one obtains the order-of-magnitude estimate: 
\begin{equation}
p_{\max} \sim\frac{v_A}{\nu_{\rm IN}} v_{\rm sh} B \frac{P_{\rm CR}}{\ln\Lambda} 
\end{equation}

Using $v_A=\frac{B}{\sqrt{4\pi \rho_i}}$,  $\rho_i=m_p n_i$,
\begin{equation}
\begin{aligned}
p_{\max}&\sim \frac{1}{\ln\Lambda} \frac{\xi\,\rho\,v_{\rm sh}^3}{\alpha\,T_4^{0.4}} \frac{B^2}{\sqrt{4\pi m_p n_i}}\\
&
\approx C \xi \left(\frac{v_{\rm sh}}{100 \text{km/s}}\right)^3 \left(\frac{T}{10^4 \text{K}} \right)^{-0.4}  \left( \frac{ B}{3 \mu \rm G}\right)^2 \frac{\sqrt{n_i}}{n_n}  \left(\frac{7}{\ln\Lambda}\right){\rm GeV}/c,
\end{aligned}
\end{equation}

and the coefficient $C$ is of order unity to a few tens depending on the adopted normalization of $\nu_{\rm IN}$.

introducing the ionization fraction $x_i=n_i/n_H$, so that $n_n = (1-x_i)n_H$, this leads to: 
\begin{equation}
p_{\max}\approx 39 \,\xi
\left(\frac{v_{\rm sh}}{100\,{\rm km/s}}\right)^3 \left(\frac{T}{10^4 {\rm K}}\right)^{-0.4} \frac{\sqrt{x_i}}{1-x_i} \frac{1}{\sqrt{n_H}} \left(\frac{7}{\ln\Lambda}\right)  \text{GeV/c},
\end{equation}
assuming $\nu_{\rm IN}\simeq 10^{-9} T_4^{0.4} n_n \,{\rm s^{-1}}$, 
and
\begin{equation}
p_{\max}\approx 4.7 \,\xi
\left(\frac{v_{\rm sh}}{100 {\rm km/s}}\right)^3
\left(\frac{T}{10^4 {\rm K}}\right)^{-0.4} \frac{\sqrt{x_i}}{1-x_i} \frac{1}{\sqrt{n_H}} \left(\frac{7}{\ln\Lambda}\right) {\rm GeV}/c
\end{equation} 

assuming $\nu_{\rm IN}\simeq 8.4 \times10^{-9} T_4^{0.4} n_n \,{\rm s^{-1}}$.
This recovers the typical "linear" scaling of~\citet{drury1996}, up to normalization factors absorbed into $C$.
In linear estimates such as~\citet{drury1996}, the CR pressure gradient is treated as an external input. This leads to a maximum energy that depends explicitly on $P_{\rm CR}$. Differently, in self-consistent models~\citep{ptuskin2003}, the diffusion coefficient and the CR gradient are coupled through the wave amplitude, thus regulating the instability. In this case, the dependence on $P_{\rm CR}$ cancels out. \citet{eichler2017} reaches a somewhat similar result by directly imposing the condition $\Gamma_{\rm CR}=\Gamma_{\rm IN}$ without solving the full transport problem.

\begin{figure}
    \centering
    \includegraphics[width=0.4\textwidth]{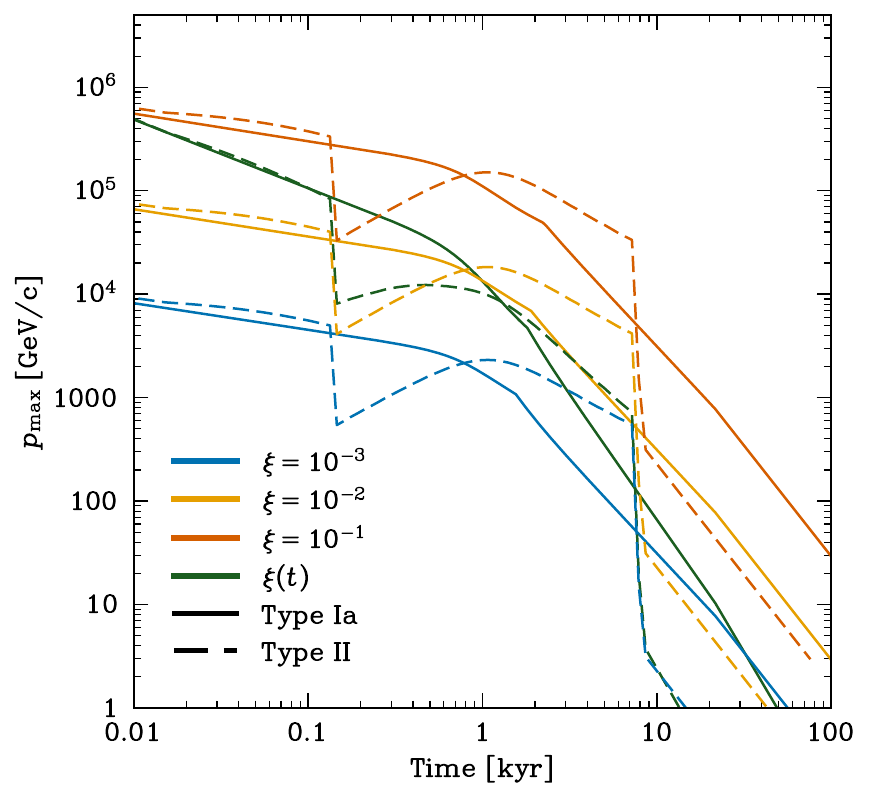}
        \includegraphics[width=0.4\textwidth]{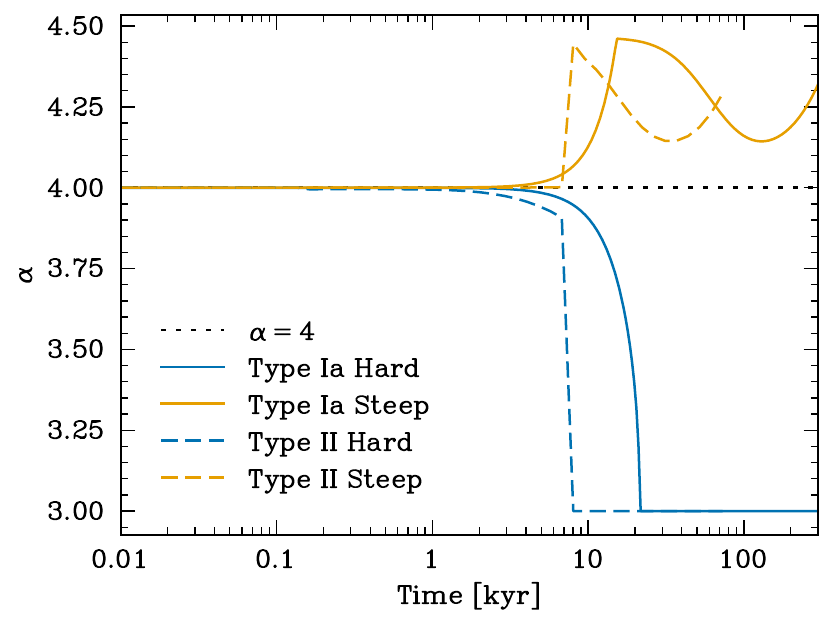}
\caption{Top panel: maximum momentum of accelerated particles for the two benchmark SNRs, Type Ia (solid lines) and Type II (dashed lines). Three constant values of $\xi$ ($0.1$, $0.01$, and $0.001$) are shown, together with the time-dependent case $\xi(t)$ discussed in the text. Bottom panel: spectral slope of accelerated particles for the Type Ia and Type II SNRs. The "hard'' case corresponds to $\alpha<4$, obtained at radiative shocks with an enhanced compression ratio; the "steep" case corresponds to $\alpha>4$, when particle acceleration is controlled by a radiative precursor with a reduced subshock compression ratio; the canonical test-particle case corresponds to $\alpha=4$}
    \label{fig:pmax}
\end{figure}

The maximum momentum of accelerated particles for the two benchmark Type Ia and Type II SNRs, with parameters adopted from~\citet{cristofari2021}, is illustrated in Fig.~\ref{fig:pmax}. We consider three cases for the spectral slope of the accelerated particles: "hard'', corresponding to $\alpha<4$ at a radiative shock with an enhanced compression ratio; "steep'', corresponding to $\alpha>4$ when particle acceleration is governed by a radiative precursor with a reduced subshock compression ratio; and the canonical case with $\alpha=4$, corresponding to the test-particle limit of DSA. The non-monotonic evolution of the spectral slope in the Type~Ia case results from the competition between radiative cooling and shock deceleration. When the shock enters the radiative phase, the increasing radiative flux enhances the upstream photoionization and heating, thus reducing the shock Mach number and the subshock compression ratio, which leads to a steeper accelerated particle spectrum. At later times, the decrease in shock velocity reduces both the radiative energy flux and the ionizing photon production, thus weakening the radiative precursor. The upstream temperature consequently decreases, allowing the subshock compression ratio to partially recover and the particle spectrum to flatten again.

\section{Results and discussion}
\begin{figure}
    \centering
    \includegraphics[width=0.48\textwidth]{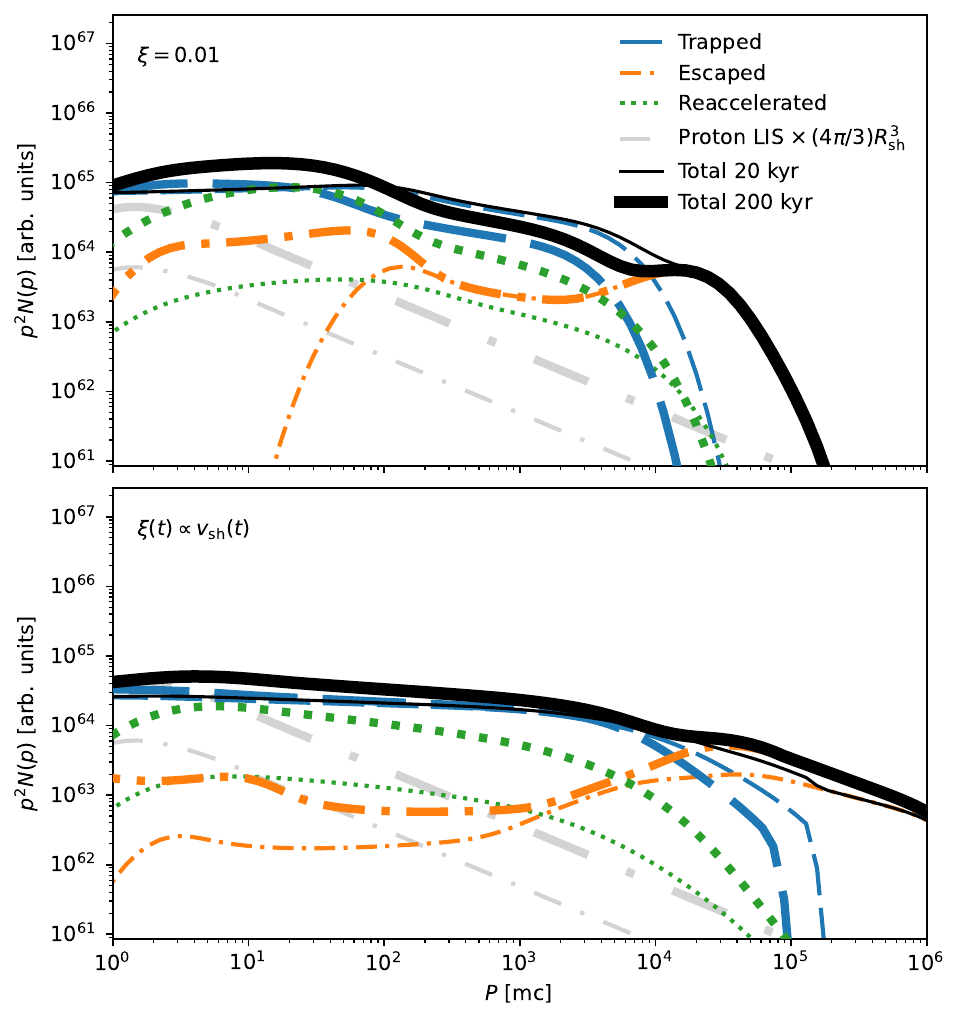}
\caption{Momentum spectra of trapped, escaped, and reaccelerated particles, together with the resulting total spectrum, for the typical end of the ST phase at 20~kyr (thin lines) and the end of the radiative phase at 200~kyr (thick lines), in the case of a Type Ia SNR. The light-gray dash-dotted lines show the proton LIS multiplied by the corresponding SNR volume, $V_{\rm SNR}=4\pi R_{\rm sh}^3/3$. The upper panel assumes a constant cosmic-ray acceleration efficiency, $\xi_{\rm CR}=0.01$, whereas the lower panel adopts a time-dependent efficiency scaling as $\xi_{\rm CR}\propto v_{\rm sh}$.}
    \label{fig:contributions}
\end{figure}

We compute the total CR proton content accelerated over the lifetime of a SNR up to the end of the radiative phase, for SNRs from prototypical Type Ia and Type II SNe.  Freshly accelerated particles, reaccelerated pre-existing cosmic rays trapped within the SNR, and particles escaping from the remnant all contribute to shaping the overall spectrum injected by the SNR into the ISM, as illustrated in Fig.~\ref{fig:contributions}. For comparison, we also show the proton LIS multiplied by the SNR volume at each age, providing a reference for the amount of pre-existing CRs swept up by the remnant and available for reacceleration.
When particle acceleration remains sufficiently efficient during the radiative phase to produce a hard spectrum of freshly accelerated particles, reaccelerated pre-existing CRs contribute at a level comparable to that of freshly accelerated particles at low energies. As a result, they play an important role in shaping the total CR spectrum injected into the ISM. In contrast, if the acceleration efficiency decreases with time, for instance as $\xi_{\rm CR}\propto v_{\rm sh}$, the contribution of freshly accelerated particles produced during the radiative phase is significantly reduced. Consequently, the relative contributions of the different particle populations are altered, leading to a different CR spectrum injected into the ISM.

The main difference between these two cases is that, for Type II events, the shock wave propagates through a circumstellar medium (CSM) structured by the life of the progenitor star. In this case, the shock successively propagates through the dense wind of the late evolutionary stages of the star, then through the more diluted wind bubble created during earlier phases, and finally into the unperturbed interstellar medium~\citep{weaver1977,ptuskin2005}. Given the large diversity of progenitor star properties -  including total mass, ejecta mass, explosion energy, mass-loss rates, and wind velocities- the resulting circumstellar medium (CSM) structure can vary significantly. In this work, we adopt representative values for these quantities, while keeping in mind the wide diversity observed in the Galaxy.

\begin{figure*}
    \centering
\begin{tabular}{cc}
\includegraphics[width=0.49\linewidth]{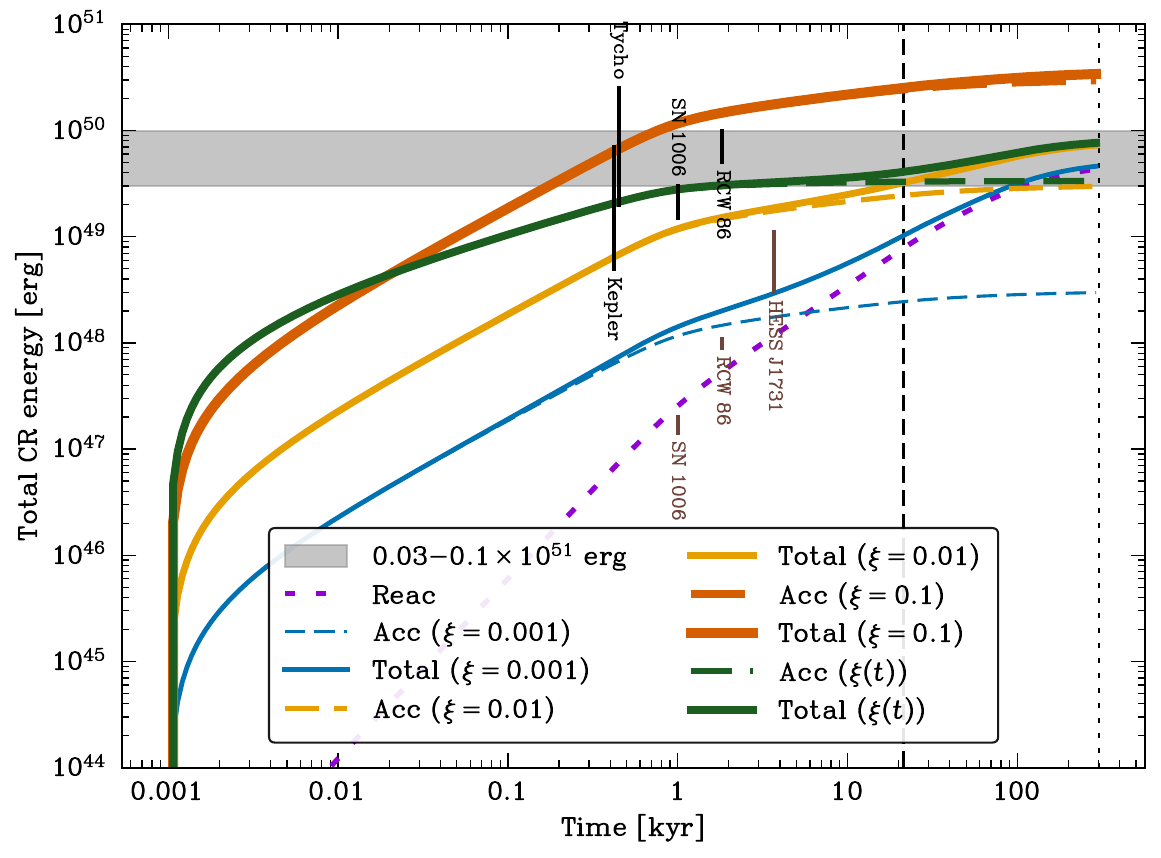} &
\includegraphics[width=0.49\linewidth]{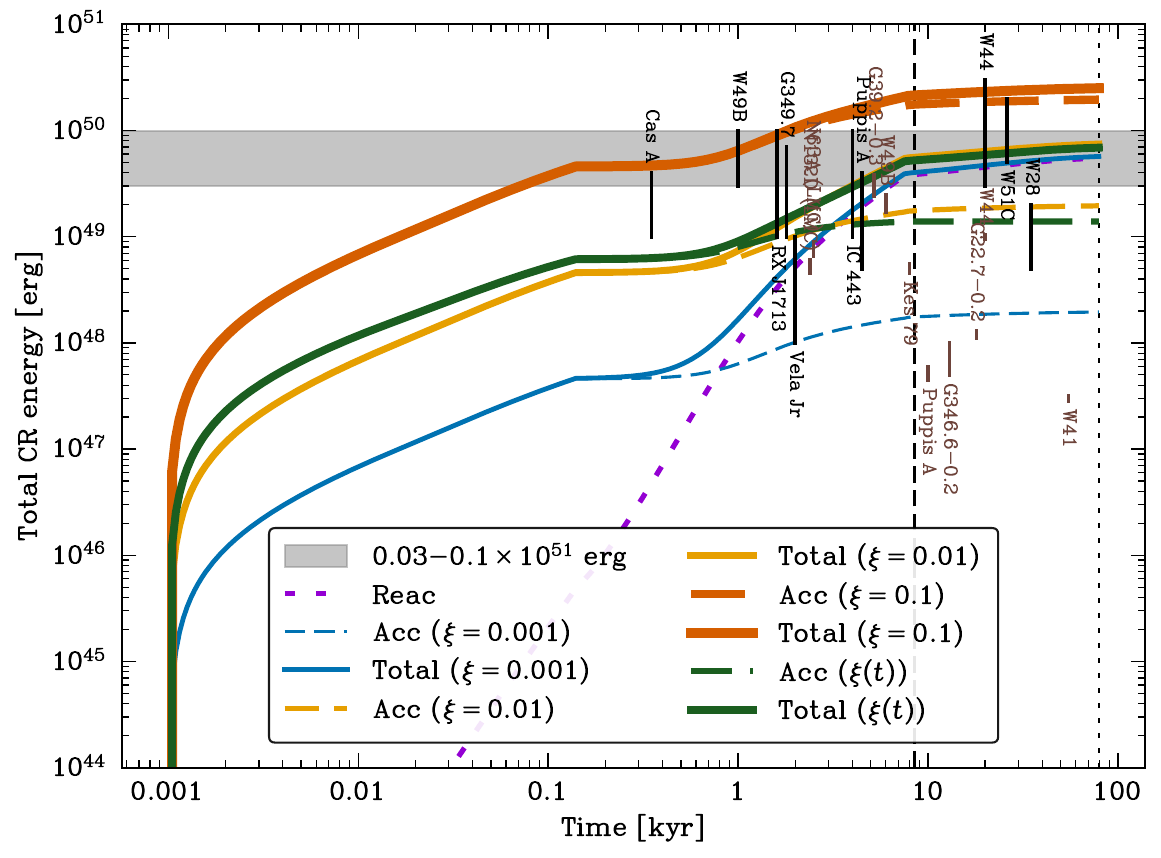} \\
\end{tabular}
    \caption{Total CR proton energy as a function of time at the shock of a Type Ia (left) and Type II (right) SNR. Accelerated protons (dashed lines) are shown for CR acceleration efficiencies of 0.1\% (blue), 1\% (orange), and 10\% (vermillion). Reaccelerated protons are shown with dotted violet lines, while the total contribution of accelerated and reaccelerated protons is shown with solid lines. The vertical dashed and dotted lines indicate the approximate end of the Sedov--Taylor (adiabatic) phase and the onset of the radiative phase, respectively. The horizontal grey area marks the  3-10\% of the canonical supernova explosion energy ($10^{51}$ erg), corresponding to the fraction typically required for supernovae to sustain the observed Galactic CR power. The estimated values for the SNRs detected in gamma rays and often proposed as remnants from Type Ia SNe, are shown in black: Kepler~\citep{aharonian2022,acero2022}, Tycho~\citep{acciari2011,archambault2017}, SN~1006~\citep{acero2010,condon2017}, and RCW~86~\citep{aharonian2009,yuan2014},  Values inferred from refined analyses exploiting high-angular resolution HI and $^{12}$CO data are shown in brown~\citep{sano2021,sano2023,ito2025,murase2026}. For Type II SNRs: Cas A~\citep{abdo2010}, W49B~\citep{W49B_HESS,sano2021}, RXJ1713~\citep{RXJ1713_HESS,sano2019}, Vela Jr~\citep{abdalla20218,sano2021}, W49B~\citep{abdalla2018b}, Puppis A~\citep{hewitt2012,xin2017}, IC443~\citep{ackermann2013}, W44~\citep{giuliani2011,cardillo2014}, Kes 79~\citep{auchettl2014}, W41~\citep{castro2013,murase2026},  N132D (LMC)~\citep{sano2020}, N63A (LMC)~\citep{sano2019}, G39.2-0.3~\citep{deonewilhelmi2020}, G346.6-0.2~\citep{sano2021}, G22.7-0.2~\citep{murase2026}.} 
    \label{fig:SNR1_energy}
\end{figure*}

Assuming an efficiency constant in time $\xi=10$ \%, the total energy in accelerated protons exceeds 5 \% (10 \%) of the canonical SN total explosion energy $10^{51}$ erg for ages $\gtrsim 1$ kyr ($\gtrsim 3-5$ kyr). This means that before the end of the ST phase, the total content of CR protons from SNR begins to exceed what is needed to account for the LIS. This supports the idea that $\xi$ decreases in time, or that, if constant in time, $\xi<10$ \%. $\xi=1$ \%,  0.1 \%, or  $\xi\propto t^{-1}$ allow to not overproduce the CR proton LIS by typically leading to $\sim 0.05- 0.1 \times 10^{51}$ erg at the end of the radiative phase (see Fig.~\ref{fig:SNR1_energy}). For reference, the typical estimate of the CR proton content derived from gamma ray observations at historical SNRs are shown. We additionally show the CR proton estimates obtained in the case of  refined analysis exploiting multi-wavelength data, and often concluding to mixed hadronic and leptonic contribution to the gamma-ray signal. 

The parameter $\xi$ affects the normalization of the freshly accelerated component, whereas the total proton energy shown in Fig.~\ref{fig:SNR1_energy} also includes reaccelerated CRs, whose contribution becomes significant at late times.  Reacceleration also acts on secondary nuclei and can modify secondary-to-primary ratios such as Boron-over-Carbon~\citep{blasi2017,bresci2019}. In order to verify if the low-$\xi$ scenarios are fully compatible with the constraints on the composition of CRs, a self-consistent Galactic transport calculation is needed, which is beyond the scope of this work.

The total cumulative proton spectra up to the end of the radiative phase leads to a deviation from $\propto p^{-4}$ up to $\sim 10$ TeV in the case of Type Ia. The integration in time between 100 kyr and 300 kyr has a minimal impact, as in this part of the radiative phase, particle acceleration is drastically limited by ion-neutral damping. The various recipes used to estimate the effective compression ratio relevant for particle acceleration shape the spectrum below $\lesssim 1$ TeV.  In the "hard" case, in which the slope of accelerated particles $\alpha <4$ in the radiative phase, the hardening appears below $10$ GeV and steeping at higher energy. The "bump" produced in the 10-100 GeV range is due to the reacceleration of preexisting CRs. The case $\alpha=4$ in which the canonical DSA test-particle value is assumed in the entire life of the SNR also leads to a steepening with a slope for the cumulative spectrum in the range 4.1 - 4.3. The "steep" case, accounting for the possibility of an effective reduced compression factor $r_{\rm sub}<4$ due to heating upstream as discussed in Sec.~\ref{sec:rad}, and thus $\alpha>4$ , produces a spectral index  of the cumulative spectrum in the range 4.2 - 4.3 below $1$ TeV. In the case of Type II, the role of the radiative phase is rather limited, and the various prescriptions for $\alpha$ only affect the cumulative spectrum below $\lesssim 100$ GeV.  

The difference between the cumulative spectrum from Type Ia and Type II SNRs motivated us to investigate the possible consequence on the helium spectrum. It has now been clearly established that from the GeV to the PeV range, the proton and helium spectra  exhibit a complex relative evolution~\citep{aguilar2015_protons,aguilar2015_helium,DAMPE2019,DAMPE2021,cao2026}, with the helium spectrum becoming alternately harder and softer than the proton spectrum, a behaviour that is not yet understood. 
In a medium of approximately solar composition, the helium abundance is expected to be of order $n_{\rm he}/n_{\rm p}\simeq 0.1$ by number~\citep{asplund2021}, which is the appropriate reference value for most SNR shocks expanding in the ISM, or in the H-rich winds of ordinary RSG progenitors. However, this abundance may be substantially
larger for some CCSNe associated with stripped-envelope progenitors. In particular, some Wolf-Rayet and stripped helium
stars can lose a substantial part of their hydrogen envelope before the SN explosion, thus leading to winds and CSM that are H-poor and He-rich~\citep{crowther2007,smith2014}. 
Direct observational evidence is for example provided by Type Ibn SNe, such as SN~2006jc, SN~2000er, SN~2002ao, SN~1999cq, SN~2010al,
and SN~2011hw, whose spectra show prominent narrow He\,{\rm I} lines and weak
or absent hydrogen features, indicating interaction with He-rich CSM
\citep{foley2007,pastorello2008,pastorello2015}.

\begin{figure}
    \centering
    \includegraphics[width=0.48\textwidth]{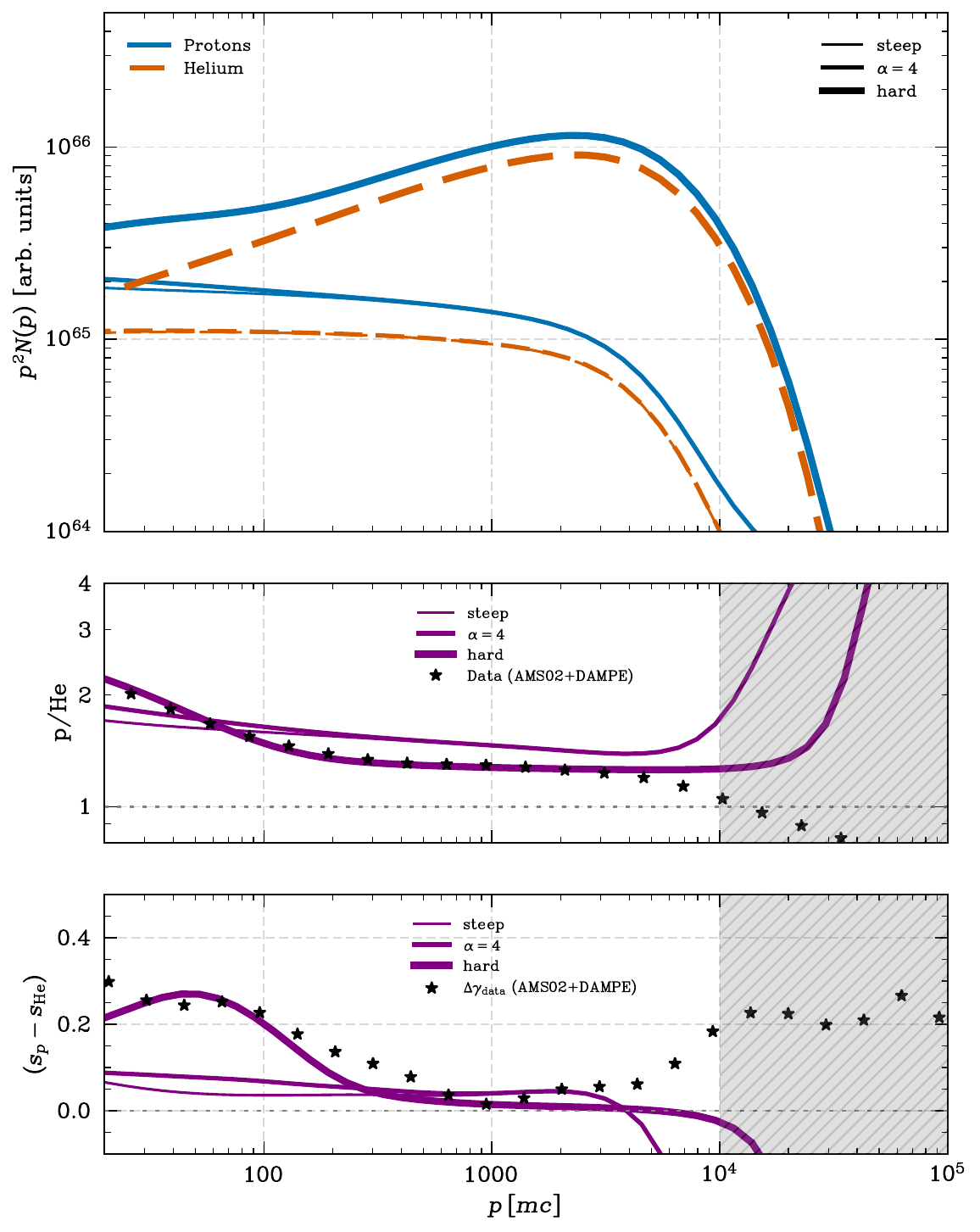}
\caption{
Top panel: Proton (solid blue lines) and helium (dashed orange lines) spectra as a function of momentum, obtained by summing the contributions from thermonuclear and core-collapse SNRs. The helium abundance around core-collapse SNRs is assumed to be a factor of 10 higher than around thermonuclear SNRs. The steep, test-particle, and hard acceleration cases are shown from thin to thick lines. Middle panel: Proton-to-helium ratio as a function of momentum. The data (black stars) are from AMS-02~\citep{aguilar2015_protons,aguilar2015_helium} and DAMPE~\citep{DAMPE2019,DAMPE2021,DAMPE2024}. The grey hatched region indicates combinations of model parameters for which no satisfactory fit to the data is obtained. Bottom panel: Corresponding difference between the proton and helium spectral slopes.}
    \label{fig:helium}
\end{figure}

This motivates us to investigate the possibility that the helium abundance is an order of magnitude larger than the typical 0.1 ratio. Here, we simply assume that at Type II SNRs, the helium-to-proton ratio is a factor of 10 larger than at Type Ia. As illustrated in Fig.~\ref{fig:helium}, the resulting difference in slope between protons and helium, together with the modified proton-to-helium ratio, can qualitatively account for the AMS-02 and DAMPE measurements below $\lesssim 10$~TeV. We restrict this comparison to this energy range, where the radiative phase can affect the cumulative SNR spectra; the behaviour at higher energies likely involves additional effects or source populations not considered here.
However, the limited mass of He-rich circumstellar material could make this enrichment mainly relevant to early acceleration and high energies; the constant abundance contrast adopted here should therefore be regarded as phenomenological. At GeV energies, the helium injection may instead be suppressed relative to hydrogen in partially ionized media because of its higher ionization potential~\citep{drury2011}. This effect was not included in our calculation.
This change in composition assumed here only affects the relative normalization of the helium and proton spectra, while $\xi$ is kept unchanged.
The fact that these helium-enriched environment could be connected to Wolf-Rayet stars is also qualitatively consistent with the long-standing interpretation of the enhanced
$^{22}$Ne/$^{20}$Ne ratio measured in Galactic cosmic rays, which has been attributed to the acceleration of material originating from Wolf-Rayet winds
or other He-rich stellar ejecta \citep{higdon2003,binns2005,tatischeff2021}.

\begin{figure*}
    \centering
    \includegraphics[width=0.42\textwidth]{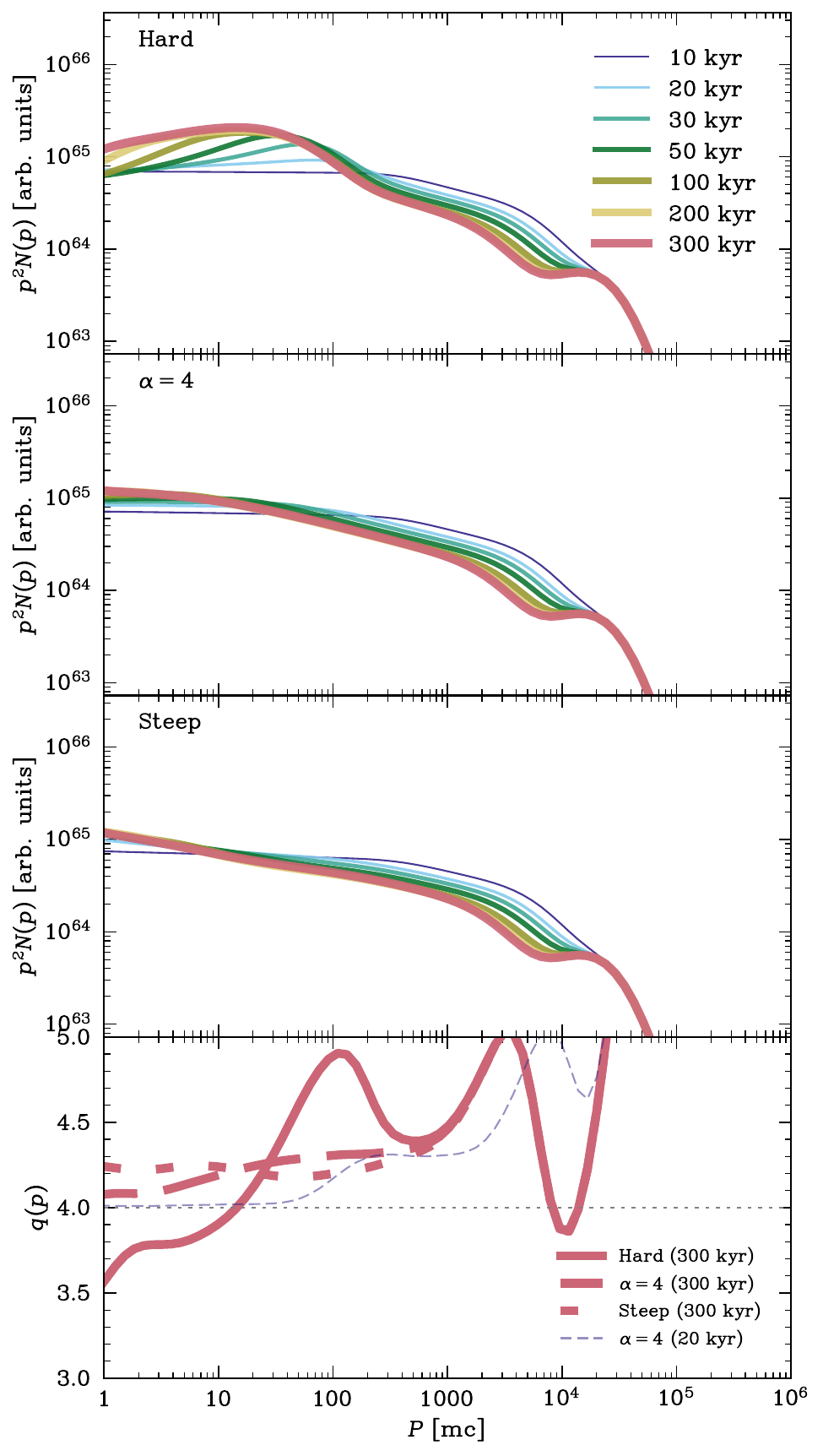}
        \includegraphics[width=0.42\textwidth]{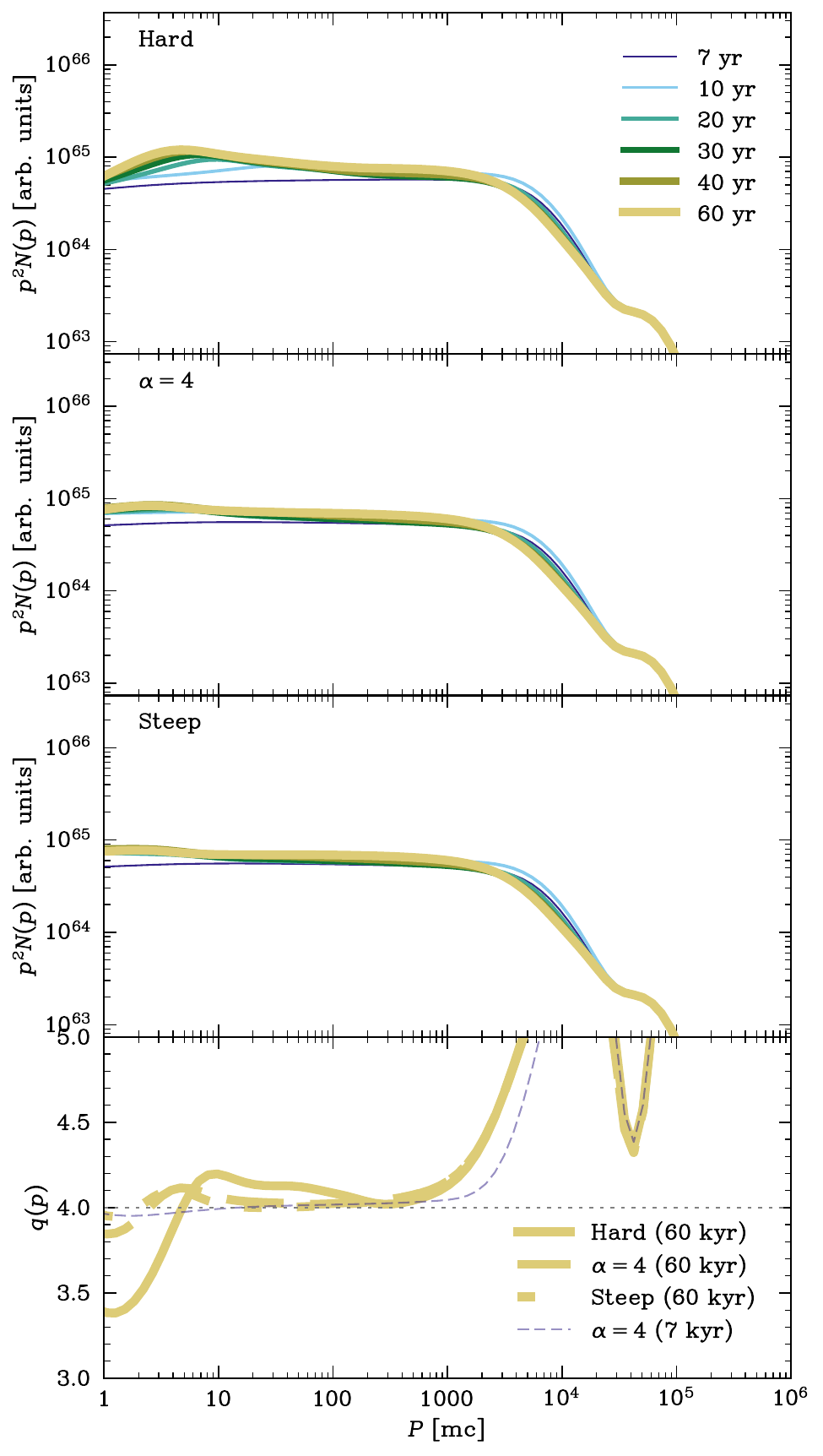}
\caption{Total CR proton spectra at a Type Ia (left) and Type II SNR (right). The hard, test-particle, and steep cases correspond to different assumptions for the slope of accelerated particles during the radiative phase. The hard spectrum is obtained for a large compression ratio, the test-particle case assumes $r=4$, and the steep case corresponds to acceleration at a subshock formed due to photoionization (see Sec.~\ref{sec:rad}).}
    \label{fig:multi}
\end{figure*}

\section{Conclusions}

We estimated the total time-integrated CR proton spectrum from typical Type Ia and Type II SNRs, taking into account part of the radiative phase during which DSA might still be efficient. We find that this phase mightplay a role in shaping the spectrum injected into the ISM up to $\sim 10$~TeV.
 Our approach is rather rudimentary, in the sense that it relies on a series of simplifying assumptions that allow us to carry the calculation analytically: spherical shock, particles confined in the SNR until release in the ISM at the end of the SNR lifetime, simple magnetic field geometry described by one average value, etc. Several of these ingredients need to be investigated into more details. Especially, the actual spectrum of accelerated protons in the radiative phase is yet not clear. 

The numerous potential contributors to the Galactic CR spectrum make it important to clarify the role played by SNRs. If several source classes contribute significantly to the CR spectrum, the required efficiency of SNRs is necessarily reduced compared to the canonical value of 10\% of the total SN explosion energy.
It is important to distinguish between the fraction of the total power (or energy) injected by SNRs into the CR population $\eta_{\rm CR}$, and the instantaneous CR efficiency $\xi$, such as the fraction of the fluid ram pressure that contributes to energizing particles. 
For most of the SNR evolution, we found that the CR efficiency $\xi$ should be closer to $\lesssim 1$ \% than the canonical $5-10 \%$ in order not to overproduce the CR proton LIS. For some of the estimates of the CR proton content based on analyses taking into account the detailed mass content from $^{12}$CO and HI measurements, proton CR efficiency could be $\xi \sim 0.1$\%, making this contribution from DSA somewhat comparable to the one expected from DSRA at ages as early as $\sim 3-4$ kyr. 
 
 Moreover, if the radiative phase of SNRs contributes, in which the maximum energy of accelerated particles is usually $\lesssim 1$ TeV - or smaller when other effects such as ion-neutral damping become important- this reduces further the value of $\xi$. This is compatible with the consensual picture in which $\xi$ decreases in time. 
 
The spectral steepening at energies $\lesssim 1$ TeV, potentially resulting from DSA operating during part of the radiative phase, appears to be more pronounced for Type Ia than for Type II SNRs. We illustrated that, under the simple assumption that the helium fraction is a factor $\sim 10$ higher around Type II SNRs, acceleration during the radiative phase could help account for the observed He-to-p ratio at energies $\lesssim 10$ TeV. This illustrates that the radiative phase may play an important role not only in shaping the CR proton and helium spectra, but also, as previously discussed for electrons, which remain confined longer within SNRs and consequently develop steeper injected spectra~\citep{cristofari2025}.

 In our approach, a reduced value of $\xi$ comes from the fact that the normalization of the total CR proton content is proportional to $\xi$ and thus to limit the contribution of SNRs to a reasonable amount of CR protons, $\xi$ needs be reduced. Let us mention that at young SNRs (free expansion and early Sedov phase) $p_{\rm max}$ is also proportional to $\xi$ when it is set by the growth of non-resonant modes~\citep{bell2004,bell2013,schure2014}. $\xi \lesssim 1 \%$ in the early evolution of SNR would imply that SNRs are even further from being efficient proton pevatrons as usually discussed~\citep{cristofari2021}. This would be in agreement with the possibility that the DAMPE 10 TeV bump is produced by the maximum accelerated particles at various SNRs~\citep{recchia2024}, and this would thus reinforce the idea that other accelerators - or a subset of unusual and rare SNRs- contribute at higher energy. One of the results of our study is that the impact of the radiative phase on the cumulative spectra  from Type Ia and Type II SNRs is essentially seen in the 1 GeV - 1 TeV range, producing a steepening or hardening in the Type Ia (depending on the slope of accelerated particles assumed in the radiative phase).

An additional  consequence of the suggested low  $\xi \lesssim 1\%$ at SNR shocks, is that non-linear effects, for example due to the copious production of CRs~\citep{malkov2001,amato2006}, or to the drift of scattering centers upstream or downstream~\citep{zirakahsvili2008,caprioli2018,cristofari2022} should then play only a minimal role, both on  the deviation from the test-particle spectrum at the shock  and on their effect on the cumulative spectrum injected in the ISM. Deviations from $p^{-4}$ to produce a spectrum injected in the ISM $\propto p^{-4.2}$ would then rather be due to the trapping, confinement and losses suffered by particles, as well as cumulative piling up during the FE, ST and a period of the radiative phase. In this picture, the duration of the active radiative phase contributing to particle acceleration, the maximum energy, and the spectrum of accelerated particles emerge as key ingredients shaping the SNR contribution to the CR spectrum, highlighting the need for more detailed studies of particle acceleration in radiative shocks.

\begin{acknowledgements}
The authors thank Pasquale Blasi for stimulating discussions and valuable feedback on this work.
PC and MVHP acknowledge support from the "GALAPAGOS" PSL Starting Grant and from the "PARMIGIANO" ANR. 
\end{acknowledgements}

%
   \bibliographystyle{aa} 
   \bibliography{radiative} 
%

\end{document}